# The Garber Current Pattern: An Additional Contribution to AC Losses in Helical HTS Cables?

Steffen Elschner, Andrej Kudymow, Nicolo Riva, Francesco Grilli

*Abstract*—Conductors made of high-temperature (HTS) wires helically wound in one or more layers on round tubes (CORT) are compact, flexible, and can carry a large amount of current. Although these conductors were initially developed for DC applications, e.g. in magnets, it is worth considering their use for AC, e.g. in underground cables for medium voltage grids and with currents in the kA-range. In these cases, the major challenge is reducing AC losses. In contrast to a straight superconducting wire, in a helical arrangement, due to superconducting shielding, the current does not follow the direction of the wires, but takes a non-trivial zig-zag path within the individual HTS wires (Garber pattern). This includes current components across the thickness of the superconducting layers, so that the often used thin-shell approximation does not hold. In this contribution, we studied a one-layer three-wire CORT by means of fully three-dimensional simulations, based on the H-formulation of Maxwell's equations implemented in the commercial software package COMSOL Multiphysics. As a result of our simulations, the peculiar current profiles were confirmed. In addition, the influence of current, pitch angle, and frequency on the AC losses was studied. We found an optimum for the pitch angle and that the current profiles strongly depend on frequency.

## I. INTRODUCTION

THE Conductor-On-Round-Tube (CORT) is a promising design for cables made of REBCO high-temperature superconducting (HTS) wires, such as CORC® [1][2].

These cables are primarily intended for high-field magnets, but their small size, flexibility and excellent current-carrying performance make them an attractive candidate for AC power applications as well. For these applications the AC loss level is an important factor affecting the practical realization. In general, the AC loss of HTS strongly depends on how the current and the magnetic field, both strongly related, distribute inside the superconductor.

Helical conductor configurations such as CORT or concentric HTS cables [3] have a particular magnetic field symmetry. For an idealized arrangement (wires on a round former, one layer, infinite length, no gaps), the magnetic field on the inner side of the helix is strictly axial, whereas on the outer side it is, according to Ampère's law, strictly azimuthal.

In the case of a normal conducting helix in DC or at low frequencies, the current can be assumed to be homogeneous and directed along the wire. From the inner to the outer side the axial field decreases and the azimuthal field increases.

If superconductors are used, additional features appear. In particular, a peculiar zig-zag current pattern and regions with radial currents. This was recognized by Garber *et al.* already in 1976 [4]; for this reason, in the following, we will call this the 'Garber current pattern' (see sect. II). This effect was later revisited by Clem and Malozemoff [5][6], in the context of the calculation of AC losses. However, until now, this effect has only been discussed within the framework of the critical state model (CSM) [7][8].

A summary of the different contributions to AC losses in CORT-type cables is given in [5]. For a one-layer CORT, two contributions are relevant. The first is the surface hysteresis losses which, within the CSM, can be calculated analytically [8][9]. The second is due to the perpendicular field components in the gaps between adjacent wires. For the latter, analytical expressions have been developed [10][11], under the approximation of infinitely thin straight wires arranged conformally to a round former.

In terms of numerical simulations for calculating the AC losses of CORT cables, in the literature there are two simplified approaches: (i) The cable and the wires are considered straight and one simulates the transversal cross-section in 2D [12]; (ii) The helicoidal winding of the wires is considered, but the thickness of the superconducting layer is neglected [13]. Neither approach is suitable for analyzing the Garber current pattern (which is a 3D effect). This work constitutes a first attempt at using a fully 3D numerical model to visualize this current pattern and investigate its influence on AC losses.

There are not many reports on experimental measurements of transport AC losses of CORT cables. In [14], the cyclic AC losses increase with the second power of the transport current, a sign of the possible influence of the eddy current losses in the copper former. In [15], losses four times higher than for a numerical model based on straight wires are reported, and the authors indicate the irregularity of the gaps and the non-uniform

This work was supported by Germany's Federal Ministry for Economic Affairs and Climate Action – Grant Nr. 03EN4031B.

Steffen Elschner (s.elschner@hs-mannheim.de, corresponding author) is with University of Applied Science Mannheim, Germany, and Karlsruhe Institute of Technology, Germany.

Nicolo Riva (nicoriva@mit.edu) is with the Plasma Science and Fusion Center, Massachusetts Institute of Technology, United States of America.

Andrej Kudymow (Andrej.Kudymow@kit.edu) and Francesco Grilli (Francesco.Grilli@kit.edu) are with Institute of Technical Physics, Karlsruhe Institute of Technology, Germany.

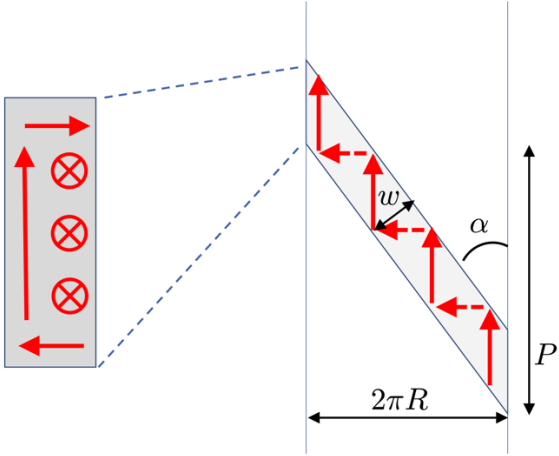

**Fig. 1.** Garber current pattern in an HTS wire. Axial current outside, azimuthal current inside. *w*: width of the wire, *α*: pitch angle, *P*: pitch length, *R*: radius of the cable.

current distribution between the tapes as possible causes. In [16], several CORT cables are considered, from one-layer cables with large gaps to multi-layer cables: Generally, it is found that losses are lower than those predicted by the Norris model for a thin strip [17] with the same critical current. This is attributed to the cancellation of the perpendicular field. In short, the experiments conducted so far do not provide a clear picture of the transport AC losses of CORT cables, which further justifies the necessity of developing accurate numerical models.

## II. THE GARBER CURRENT PATTERN

We consider an idealized one-layer CORT with a pitch angle *α*. In the CSM approximation and at low and moderate currents, a shielded neutral zone in the inner of the wire must be postulated. With Ampère's law, this neutral zone is only obtained if the axial field is shielded by a strictly azimuthal current on the inner side of the conductor and the azimuthal field by an axial current on the outer side. Together with the imposed transport current *I(t)*, this leads to a peculiar zig-zag current pattern: on the outer side of the helix, the current path is axial. When the current reaches the edge of the wire it turns towards the inner side of the helix, i.e., it follows a path along the radius of the helix, perpendicular to the *ab*-planes of the superconductor. On the inner side of the helix, the current then follows the azimuthal direction along the circumference until it reaches again an edge and turns up to the outer surface of the conductor. In the following, we call the region near the edges with the current perpendicular to the wire plane 'diving zone'. Fig. 1 shows a sketch of the current path in the HTS wire, plotted in an unrolled turn (right, see also [5][6]) and in a cross section (left).

Within the CSM, the Garber pattern has a direct impact on the critical current of the CORT. Due to the elongation of the current paths, the critical current is reduced. Straightforward geometrical considerations yield a current threshold

$$I_t = \frac{I_{c0}}{\sin\alpha + \cos\alpha} \quad (1)$$

where $I_{c0}$ is the critical current of the straight wire. At this threshold, the neutral zone disappears, and, within CSM, the critical current is reached. In the limits $\alpha = 0°$ (straight wires) and $\alpha = 90°$ (closed rings), the Garber pattern vanishes and in (1) the initial value $I_{c0}$ is recovered. For larger currents ($I > I_t$), the perpendicularly oriented flux fronts from both sides interfere (flux cutting threshold [18][19]) and the simple CSM-based description used so far loses its validity.

## III. CABLE'S GEOMETRY

In order to analyze the influence of the Garber current pattern on the critical current and the AC losses by numerical simulations, we chose the simple geometry of a one-layer CORT consisting of three wires with a width *w* = 4 mm, and a pitch angle *α* (Fig. 2) wound conforming to a cylinder. With a gap width *g* (set equal to 100 μm) the radius *R* of the CORT is

$$R = \frac{3 \cdot (w+g)}{2 \cdot \pi \cdot \cos\alpha}. \quad (2)$$

We further assume that the straight HTS wires have a superconducting thickness of 1μm, a critical current density $J_c = 4 \cdot 10^{10}$ A m$^{-2}$ and thus a critical current $I_{c0} = 160$ A and that the material is characterized by the power law:

$$E(J) = E_c \cdot \left(\frac{J}{J_c}\right)^n \quad (3)$$

where $J_c$ is the current density at $E_c = 10^{-4}$ V/m, here assumed to be independent of the magnetic field. In this first attempt, we also assume an isotropic critical current density. The parameter *n* indicates the steepness of the superconducting transition, here *n* = 25. In the limit $n \to \infty$, the power-law model converges toward the CSM. In our analysis, we only consider the superconducting layer.

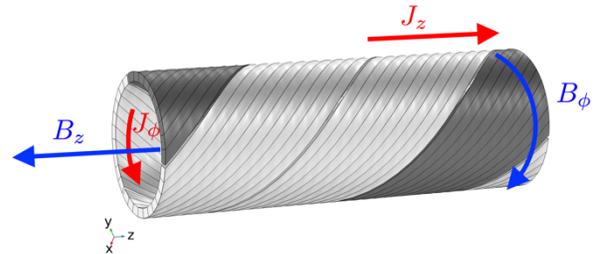

**Fig. 2.** One-layer CORT model: 3 parallel wires.

For this work, we use a full three-dimensional finite-element model, based on the well-established H-formulation of Maxwell's equations [20][21], implemented in the commercial software package COMSOL Multiphysics. Each wire is divided into ten parts (Fig. 2), in order to have more flexibility with the mesh and with the calculation of the different AC loss components. In radial direction the superconductor is divided into 20 elements, in azimuthal direction into 36 elements concentrated at the edges. The simulations are performed for one pitch length and periodic boundary conditions are imposed at the two ends of the simulated domain.

In the simulations, the thickness of the superconducting layer is artificially expanded by a factor 400 and set equal to 400 μm, in order to keep the size and the computation time of the numerical problem at an acceptable level. The critical current density is reduced accordingly ($10^8$ A m$^{-2}$), in order to keep the same current values.

The AC losses per cycle are calculated by integrating over the whole 3D-volume (three wires) and one period $T$:

$$Q = \int_T \int_V \vec{E} \cdot \vec{J} \cdot dV \, dt \qquad (4)$$

## IV. SIMULATION OF CURRENT PROFILES

With the assumptions and techniques described above, the current profiles were calculated. As an example, Fig. 3 shows the results for $J_z$ (axial), $J_\varphi = \sqrt{J_x^2 + J_y^2}$ (azimuthal) and $J_r$ (radial) in a cross-section perpendicular to the cable axis and a pitch angle $\alpha = 45°$. A peak current of 3 x 90 A$_p$ is chosen, i.e. the current remains below the threshold given in (1), in our case $I_t = 113$ A per wire. The frequency is 50 Hz. All current densities are normalized with respect to the critical current density $J_c$. A plot of the absolute value of the current density (Fig. 3, 1st from left) shows that a neutral zone is left. As expected, the current is strictly axial outside (2nd), strictly azimuthal inside (3rd), and radial at the edges (4th). The signs of $J_\phi$ and $J_r$ are related to the chosen sense of rotation (Fig. 2). For isotropic critical current density, the width of the diving zones is, as expected, in the order of the thickness of the superconducting layer (400 μm in our simulations). Thus, the electromagnetic simulations confirm the existence of the Garber current pattern including the radial current components.

## V. AC LOSSES

For the purpose of AC loss evaluation, as a first step we consider the losses caused by the longitudinal ($J_z$) and azimuthal ($J_\phi$) components of the current (see Fig. 3). The losses caused by the radial current ($J_r$) are ignored because they occur only on a very small volume which is not easy to analyze in the 3D-model. The gap losses due the perpendicular magnetic field component related to the gaps between the wires are also ignored, because not directly linked to the Garber current pattern, which would exist also for the ideal situation of isolated wires with zero gap.

Within the CSM model, the AC surface losses can be evaluated as was already done for a 2-layer CORT-cable [5]. According to the CSM, the losses per surface and cycle are then given by (see [9], eq. (3.11)):

$$\frac{Q}{A} = \frac{2B^3}{3\mu_0^2 J_c} \qquad (5)$$

However, this equation only holds as long as a neutral zone exists, i.e., until both flux fronts (with perpendicular field directions) are separated. Once the fronts meet in the inner of the superconductor, vortices of different direction intersect ('flux cutting' [5][18][19]) and the magnetic field ceases to be purely axial or azimuthal.

For small gaps the magnetic fields penetrating from outside and inside are

$$B^{out} = \frac{\mu_0 \cdot I}{2\pi R} = \frac{\mu_0 \cdot I \cdot \cos\alpha}{3(w+g)} \qquad (6)$$

and, with the periodicity (pitch) length $P$,

$$B^{in} = \frac{\mu_0 \cdot I}{P} = \frac{\mu_0 \cdot I \cdot \tan\alpha}{2\pi R} = \frac{\mu_0 \cdot I \cdot \sin\alpha}{3(w+g)} \qquad (7)$$

Inserting (6) and (7) in (5) and introducing the cable length $L$ we obtain

$$\frac{Q}{L} = \frac{2\mu_0 I^3}{27 J_c (w+g)^2} \cdot \left(\frac{\cos^3\alpha + \sin^3\alpha}{\cos\alpha}\right) = \frac{Q_S}{L} \cdot \left(\frac{\cos^3\alpha + \sin^3\alpha}{\cos\alpha}\right) \qquad (8)$$

$Q_S$ are the losses of a cable with 3 straight wires conformal to a round former ($\alpha \to 0$). In the limit $g \to 0$ the well-known Norris result [17] is obtained. Within the CSM, (8) is expected to be correct for currents up to the threshold given by (1).

In order to estimate only the surface losses, the integral of (4) is performed not in the whole wire's volume, but only in the central eight (of total ten) subdivisions of the wire (see Fig. 2). The result is then multiplied by 10/8 and divided by 400. The division by 400 is necessary because otherwise the AC loss density would be integrated on a volume 400 times too large – see also (5). Somewhat surprisingly, (8) yields an angle with minimum AC losses, i.e. $\alpha = 30.94°$. However, the effect is small and the minimum is broad. In addition, according to (8), the losses at $\alpha = 0°$ and $\alpha = 45°$ are equal. For angles $\alpha \to 90°$ the losses diverge together with the radius of the CORT, (2), and thus the length of the wires.

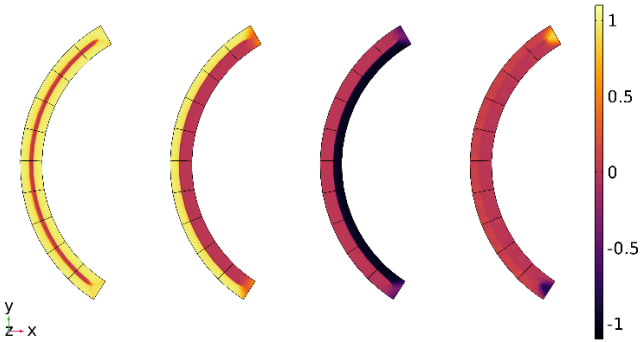

**Fig. 3.** Current density (absolute value and different components) in the cross section of one wire in the CORT (3 wires, $\alpha = 45°$, $I = 3$ x 90 A$_p$, $f = 50$ Hz).
Left to right: $J_{norm}/J_c$, $J_z/J_c$, $J_\phi/J_c$, $J_r/J_c$

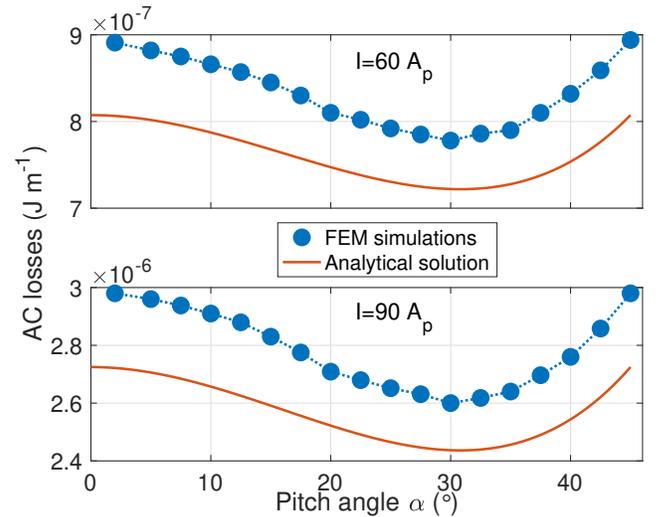

**Fig. 4.** Surface AC losses per cycle and cable length in a CORT (3 wires, $f = 50$ Hz) in dependence of the pitch angle, for two current values (60 A$_p$ and 90 A$_p$, per wire).

Fig. 4 compares the simulations of the losses and the analytic result (8) for two different currents (60 $A_p$ and 90 $A_p$). The minimum is confirmed. The discrepancy between simulations and analytically calculated values is in the order of 10 % and can be attributed to finite $n$ exponent used in the simulations, which leads to an increase of the losses compared to the CSM [22]. Additional 2D-simulations for straight wires ($\alpha = 0$) confirm this loss reduction with increasing $n$.

## VI. RELAXATION EFFECTS

Beyond CSM, an additional feature of the Garber pattern in superconducting helices appears. If we assume a power-law resistivity for the superconductor according to (3), the electrical field is different from zero at all currents. This remaining resistivity induces a relaxation of the shielding currents and a diffusion of magnetic field into the neutral zone. This relaxation effect has already been demonstrated by FEM simulations on coated conductors and strongly influences the frequency dependence of their AC losses [23][24][25].

Fig. 5 depicts the current densities in axial and azimuthal direction along the radius in the middle of the tape at different frequencies. The peak current per wire is $I_p = I_{c0} = 160$ $A_p$. At 50 Hz the Garber pattern is still well visible in the simulations (Fig. 5, straight lines), i.e. large axial currents at the outer, large azimuthal currents at the inner surface of the helix.

At 0.5 Hz (dotted lines) the current is nearly fully relaxed to a homogeneous current profile. With the pitch angle $\alpha = 45°$ both current components are equal, with a peak value $J_z = J_\phi = J_c / \sqrt{2}$, which confirms a current parallel to the direction of the wire. Thus, at high frequencies, although in the flux cutting regime, the conductor shows remainders of the CSM, the time for considerable relaxation being too short. At low frequencies, the conductor oscillates between almost fully relaxed states.

## VII. CONCLUSION AND OUTLOOK

Fully 3D numerical simulations of current profiles in helically wound HTS wires have confirmed the presence of peculiar zig-zag current profiles, including currents perpendicular to the wire's surface near the wire's edges. Below the flux cutting threshold, the one-layer CORT is well described by CSM with losses per cycle independent of frequency. At larger currents the simulations show that these current profiles are strongly frequency dependent, i.e. relaxation processes become visible, in particular at low frequencies.

This work considered only the surface losses in the superconductor. An exhaustive estimation of the AC losses of a CORT requires further study, accounting for the following effects:

1. The influence of the 'gap losses' due to the perpendicular component of the magnetic field (radial direction.) – they become important as the gap between the wires increases.
2. Strongly related, the influence of the radial currents (see Fig. 3, right), which create an additional magnetic field component in the gap between the wires (but in the direction of the wires).
3. The influence of relaxation at low frequency.
4. The critical current density was assumed to be isotropic. However, anisotropy factors up to $\gamma = J_{ab} / J_c = 600$ have been reported for the critical current densities within and perpendicular to the crystallographic $ab$ planes [26].

For the calculation of the total AC losses, one issue is that the different AC loss components scale differently with the thickness of the superconductor. In the case of the surface losses analyzed in this work, one can artificially expand the thickness of the superconductor, reduce its critical current accordingly, and reduce the AC losses calculated with (4) by the same factor (in (5) $J_c$ is at the denominator). In case of the other loss contributions such as the gap losses, this is not possible, and separating the different loss contribution is not trivial. In view of all this, it is necessary to opt for the simulation of the wire with its real dimensions, the computational cost notwithstanding. For this purpose, an approach that considers the 3D effects in a 2D simulation environment by using coordinates adapted to the symmetry of the problem (such as in [27][28]), would probably be an effective choice.

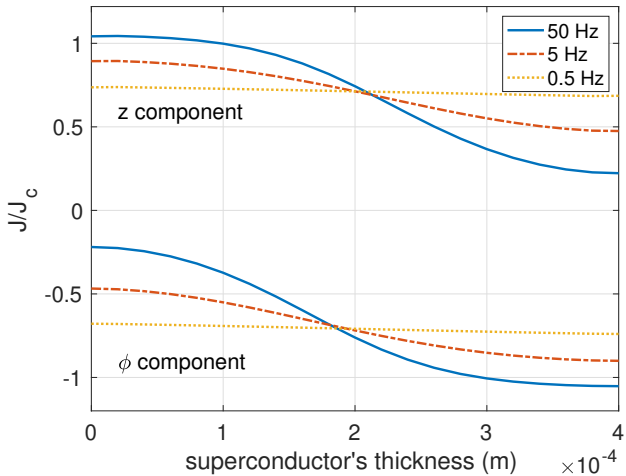

**Fig. 5.** Peak current densities $J_z / J_c$ and $J_\phi / J_c$ along the superconductor's thickness in the middle part of a wire at different AC frequencies (CORT, $\alpha = 45°$, $I = 3 \times 160$ $A_p$). The thickness is measured from the outer surface.